\documentclass[twocolumn,prl]{revtex4-1}
\usepackage[dvipdfmx]{graphicx}
\usepackage{amssymb}
\usepackage{amsmath}
\usepackage{bm}
\usepackage{color}
\usepackage{mypack}

\begin{document}

\title{Valley-dependent spin transport in monolayer transition-metal dichalcogenides}
\author{Yuya Ominato${}^1$, Junji Fujimoto${}^1$, and Mamoru Matsuo${}^{1,2}$}
\affiliation{${}^1$Kavli Institute for Theoretical Sciences, University of Chinese Academy of Sciences, Beijing 100190, China}
\affiliation{${}^2$CAS Center for Excellence in Topological Quantum Computation, University of Chinese Academy of Sciences, Beijing 100190, China}
\date{\today}

\begin{abstract}
We study valley-dependent spin transport theoretically in monolayer transition-metal dichalcogenides in which a variety of spin and valley physics are expected because of spin-valley coupling.
The results show that the spins are valley-selectively excited with appropriate carrier doping and valley polarized spin current (VPSC) is generated.
The VPSC leads to the {\it spin-current Hall effect}, transverse spin accumulation originating from the Berry curvature in momentum space.
The results indicate that spin excitations with spin-valley coupling lead to both valley and spin transport, which is promising for future low-consumption nanodevice applications.
\end{abstract}
\maketitle

{\it Introduction.---}Monolayer transition-metal dichalcogenides (TMDCs) have attracted significant attention because of their unique band structure labeled by spin and valley degrees of freedom.
Monolayer TMDCs are direct-bandgap semiconductors, and the band extrema are located at the $K_+$ and $K_-$ points of the Brillouin zone
\cite{makAtomicallyThinMoS2010,splendianiEmergingPhotoluminescenceMonolayer2010}.
Strong spin-orbit coupling of transition metals and the inversion-asymmetric crystal structure lead to spin-valley coupling (SVC)
\cite{xiaoCoupledSpinValley2012}.
The broken inversion symmetry also leads to the valley-contrasting Berry curvature \cite{xiaoValleyContrastingPhysicsGraphene2007,yaoValleydependentOptoelectronicsInversion2008,xiaoBerryPhaseEffects2010,koshinoAnomalousOrbitalMagnetism2010a,koshinoChiralOrbitalCurrent2011}, which is vitally important to assign an intrinsic magnetic moment to each valley and access the valley degrees of freedom.

Recent rapid progress in TMDC device fabrication techniques has enriched our knowledge of the valley physics, such as valley-dependent circular dichroism
\cite{caoValleyselectiveCircularDichroism2012,makControlValleyPolarization2012,zengValleyPolarizationMoS2012,sallenRobustOpticalEmission2012,wuElectricalTuningValley2013,zhaoEnhancedValleySplitting2017,zhongVanWaalsEngineering2017,seylerValleyManipulationOptically2018,nordenGiantValleySplitting2019}, the valley Hall effect \cite{makValleyHallEffect2014,leeElectricalControlValley2016a,ubrigMicroscopicOriginValley2017,wuIntrinsicValleyHall2019,hungDirectObservationValleycoupled2019a}, and valley-dependent spin injection by spin-polarized charge injection \cite{yeElectricalGenerationControl2016}.
All these experiments used charge excitations by an electric field and an optical irradiation.
Conversely, SVC provides a possible way to access the valley degrees of freedom via a spin excitation.
However, neither an experimental signature nor a theoretical proposal of spin-valley coupled phenomena by a spin excitation is missing so far.

\begin{figure}
\begin{center}
\includegraphics[width=1\hsize]{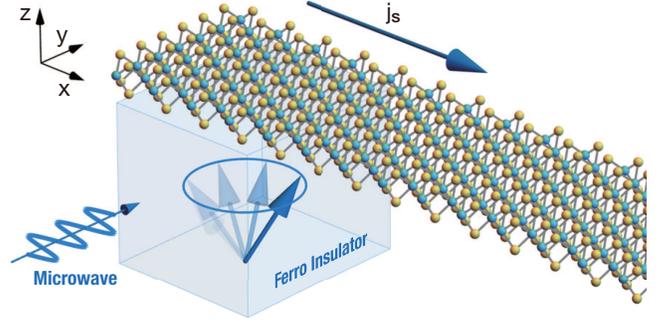}
\end{center}
\caption{A ferromagnetic insulator is fixed to the TMDC monolayer and a diffusive spin current $\bm{j}_s$ is generated by an external microwave irradiation.}
\label{fig_system}
\end{figure}

In this work, we study valley-dependent spin transport theoretically by a spin excitation in a TMDC monolayer.
Figure \ref{fig_system} shows a schematic picture of a system, in which a ferromagnetic insulator (FI) is fixed to a TMDC monolayer.
We then consider microwave irradiation of the system, which induces precession of the localized spins in the FI (i.e., ferromagnetic resonance).
The ferromagnetic resonance excites the electron spins in the TMDC monolayer via spin-transfer processes originating from the proximity exchange coupling at the interface.
We find that SVC with proximity exchange coupling leads to valley-dependent spin excitation, producing valley-polarized spin current (VPSC).
Because of the valley-contrasting Berry curvature, VPSC leads to transverse spin accumulation, which we call the {\it spin-current Hall effect}.
Solving the spin diffusion equation for the valley-polarized spins, we show the spatial distribution of the transverse spin accumulation.

{\it Model Hamiltonian.---}We consider a TMDC/FI heterostructure, where the FI weakly perturbs the band structure of the TMDC and the energy bands of the FI are absent in the energy region considered here.
The total Hamiltonian is
\begin{align}
    H=H_{\rm{TM}}+\hfi+\hex.
    \label{eq_Hamiltonian}
\end{align}
The first term $H_{\rm{TM}}=\sum_{\alpha,\bk}\e_{\alpha\bk} c^\dagger_{\alpha\bk} c_{\alpha\bk}$ describes the electronic states of the TMDC monolayer,
where $c^\dagger_{\alpha\bk}$ ($c_{\alpha\bk}$) is the electron creation (annihilation) operator with eigenenergy $\e_{\alpha\bk}$ and quantum number $\alpha=(n,\tau,s)$, where $n=\pm$, $\tau=\pm$, and $s=\pm$ are the band, valley, and spin indices, respectively.
The eigenenergy and eigenstates are derived by diagonalizing the effective Hamiltonian around the $K_+$ and $K_-$ points
\cite{xiaoCoupledSpinValley2012}:
\begin{align}
    H_{\rm{eff}}=
        &\hbar v
        \left(
            \tau k_x\s^x+
            k_y\s^y
        \right)
        +\frac{\Delta}{2}\sigma^z
        -\tau s\lambda\frac{\sigma^z-1}{2},
    \label{eq_effective}
\end{align}
where $v$ is the velocity,
$\Delta$ is the energy gap,
$\lambda$ is the spin splitting at the valence-band top caused by spin-orbit coupling, and $\bm{\sigma}$ contains the Pauli matrices acting on the orbital degrees of freedom.
These parameters are fit from first-principles calculations
\cite{zhuGiantSpinorbitinducedSpin2011,cheiwchanchamnangijQuasiparticleBandStructure2012,kormanyosMonolayerMoSTrigonal2013,liuThreebandTightbindingModel2013,kangBandOffsetsHeterostructures2013a,kormanyosTheoryTwodimensionalTransition2015,echeverrySplittingBrightDark2016}.

The second term $\hfi$ in Eq.\ (\ref{eq_Hamiltonian}) describes a bulk FI exposed to microwave irradiation:
\begin{align}
\hfi(t)=
    &\sum_\bk\hbar\omega_\bk b^\dagger_{\bk}b_{\bk}
    -\hac^+(t)b^\dagger_{\bk=\bm{0}}-\hac^-(t)b_{\bk=\bm{0}},
\end{align}
where $b^\dagger_{\bk}$ ($b_{\bk}$) is the creation (annihilation) operator for magnons with momentum $\bk$,
$\hbar\omega_\bk=Dk^2-\hbar\gamma B$ is the magnon dispersion,
and $\hac^\pm(t)=\sqrt{SN}{\hbar\gamma\hac}e^{\mp i\Omega t}/\sqrt{2}$, where
$N$ is the number of spins in the FI,
$S$ is the magnitude of the localized spin,
$B$ is a static external magnetic field,
$\hac$ and $\Omega$ are the amplitude and frequency of the microwave, respectively,
and $\gamma(<0)$ is the gyromagnetic ratio.
We have introduced in this Hamiltonian the spin-wave approximation $S^z_\bk=S-b^\dagger_\bk b_\bk$, $S_{\bk}^+=\sqrt{2S}b_{\bk}$, and $S_{-\bk}^{-}=\sqrt{2S}b_{\bk}^\dagger$, where $S^z_\bk$ and $S^\pm_\bk$ give the Fourier components of the $z$ component and of the spin-flip operators of the localized spin in the FI, respectively.

\begin{figure}
\begin{center}
\includegraphics[width=1\hsize]{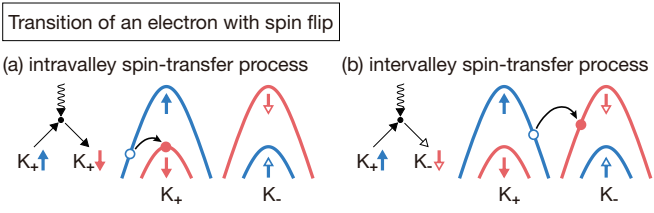}
\end{center}
\caption{The left diagrams represent (a) intravalley and (b) intervalley spin-transfer processes with magnon absorption. The solid and wavy arrows represent electrons and magnons, respectively. The right figures show schematic band structure and transition of an electron with spin flip.}
\label{fig_spin_transfer}
\end{figure}

The third term $\hex$ in Eq.\ (\ref{eq_Hamiltonian}) describes the proximity exchange coupling at the interface between the TMDC and the FI.
The proximity exchange coupling Hamiltonian contains Zeeman-like exchange coupling
\cite{qiGiantTunableValley2015a,zhangLargeSpinValleyPolarization2016,liangMagneticProximityEffect2017,xuLargeValleySplitting2018,habeAnomalousHallEffect2017,cortesTunableSpinPolarizedEdge2019}
and a tunneling Hamiltonian
\cite{ohnumaEnhancedDcSpin2014,ohnumaTheorySpinPeltier2017,matsuoSpinCurrentNoise2018,katoMicroscopicTheorySpin2019},
$\hex=H_{\rm{Z}}+H_{\rm{T}}$, where $H_{\rm{Z}}=-JSs^z_{\rm{tot}}$, and
\begin{align}
    &H_{\rm{T}}
    =-\sum_{\bq,\bk}
    \left(
        J_{\bq,\bk}s_\bq^+S_\bk^-
        +{\rm{H.c.}}
    \right),
    \label{eq_spin_transfer}
\end{align}
where $J$ and $J_{\bq,\bk}$ are the exchange-coupling constant and the matrix element for spin-transfer processes, respectively.
$s^z_{\rm{tot}}=\sum_{\alpha,\bk}sc^\dagger_{\alpha\bk}c_{\alpha\bk}$ is the $z$ component of the total electron spin in the TMDC, and $s^\pm_\bq$ is the Fourier transform of the spin-flip operators of electron spin density on the TMDC.
The Hamiltonians $H_{\rm{Z}}$ and $H_{\rm{T}}$ correspond to the out-of-plane component and the in-plane component of the proximity exchange coupling:
$H_{\rm{Z}}$ modulates the spin splitting and
$H_{\rm{T}}$ describes spin transfer at the interface, consisting of intravalley and intervalley spin-transfer processes as shown in Fig. \ref{fig_spin_transfer}.

{\it Spin current at the interface.---}The microwave excites magnons and increases magnon population, which excites spins in the TMDC monolayer because of the spin-transfer term $H_T$.
This mechanism is called spin pumping, which gives successful spin injection in bilayer systems composed of TMDCs and ferromagnets \cite{mendesEfficientSpinCharge2018,husainSpinPumpingHeusler2018,bansalExtrinsicSpinorbitCoupling2019}.
A spin excitation is described by the spin current at the interface. The spin current operator is
\begin{align}
    \IS:=-\frac{\hbar}{2}\dot{s}^z_{\rm{tot}}
        =-i\sum_{\bq,\bk}
		\left(
			J_{\bq,\bk}
			s_\bq^+
			S_\bk^-
			-
			\rm{H.c.}
		\right),
\end{align}
where we define positive current to flow from the TMDC to the FI.
We calculate the statistical average of the spin current at the interface and treat $H_{\rm{T}}$ as a perturbation and $H_{\rm{TM}}+\hfi+H_{\rm{Z}}$ as an unperturbed Hamiltonian.
The second-order perturbation calculation with respect to $H_{\rm{T}}$ gives the statistical average of the spin current at the interface
\cite{ohnumaEnhancedDcSpin2014,ohnumaTheorySpinPeltier2017,matsuoSpinCurrentNoise2018,katoMicroscopicTheorySpin2019}:
\begin{align}
    \la\IS\ra
    &=
		2\hbar
		\sum_{\bq,\bk}
		|J_{\bq,\bk}|^2
		\int_{-\infty}^\infty
		\frac{d\omega}{2\pi}
		{\rm{Im}}\chi^R_{\bq}(\omega)
        {\rm{Im}}\left[\d G_{-\bk}^<(\omega)\right].
\end{align}
The dynamical spin susceptibility of the TMDC monolayer is
\begin{align}
    &\chi^R_\bq(\omega):=
        \int^\infty_{-\infty}dte^{i\omega t}
        \frac{1}{i\hbar}\theta(t)
        \la[s_{\bq}^+(t),s_{-\bq}^-(0)]\ra.
\end{align}
The second-order perturbation calculation of the magnon propagator
$G^<_\bk(\omega):=\int^\infty_{-\infty}dte^{i\omega t}({2S}/{i\hbar})\la b_{\bk}^\dagger(0) b_{\bk}(t) \ra$
with respect to $h_{\rm{ac}}$ leads to
\begin{align}
    {\rm{Im}}\left[\d G_{-\bk}^<(\omega)\right]
    &=-\frac{1}{\hbar}
        g(\omega)
        \d(\omega-\Omega)
        \d_{\bk,{\bm 0}},
\end{align}
where $g(\omega)={2\pi N(S\gamma\hac)^2}/\left[{(\omega-\omega_{\bk=\bm{0}})^2+\ag^2\omega^2}\right]$ is the dimensionless function with the phenomenological dimensionless damping parameter $\ag$ \cite{kasuyaRelaxationMechanismsFerromagnetic1961,cherepanovSagaYIGSpectra1993,jinTemperatureDependenceSpinwave2019}.
In the current setup, only the uniform magnon mode is excited as indicated by the Kronecker delta $\d_{\bk,\bm{0}}$.

We consider an interface characterized by a roughness length scale $r$, satisfying a condition, with $k_{\rm{F}} \ll r^{-1} \ll a^{-1}$, where $k_{\rm{F}}$ is the Fermi wavelength and $a$ is the lattice constant of the TMDC.
This condition expresses an atomically flat heterostructure where (i) the matrix element is constant, $J_{\bq,\bk}=J_0$, because of the long-wavelength approximation, and (ii) the intervalley spin-transfer processes are negligible; in other wards, the roughness condition excludes a change in wave vector comparable to $a^{-1}$.

Given these conditions, we obtain the following analytical expression for the spin current:
\begin{align}
    \la\IS\ra=I_S^{K_+}+I_S^{K_-},
\end{align}
where we introduce the valley-resolved spin current
\begin{align}
    I_S^{K_\tau}=2|J_0|^2 g(\Omega){\rm{Im}}\chi^R_{\tau,{\rm{loc}}}(\Omega),
    \label{eq_spin_current}
\end{align}
with the local spin susceptibility for each valley
$\chi^R_{\tau,{\rm{loc}}}(\omega):=\sum_\bq\chi^R_{\tau,{\bq}}(\omega)$.
The imaginary part of the local spin susceptibility is given by
\begin{align}
    &{\rm{Im}}\chi_{\tau,{\rm{loc}}}^R(\omega)
        =-{2\pi\ho}
            \int d\e
            \left(
                -\frac{\partial f(\e)}{\partial \e}
            \right)
            D_{\tau,+}(\e)D_{\tau,-}(\e) \notag \\
        &\hspace{35mm}\times
        \left(
            1+\frac{Z_{\tau,+}Z_{\tau,-}}{|\e-E_{\tau,+}||\e-E_{\tau,-}|}
        \right),
        \label{eq_susceptibility}
\end{align}
where $f(\e)=1/\left(e^{(\e-\mu)/\kbt}+1\right)$ is the Fermi distribution function with chemical potential $\mu$ and temperature $T$, and
$D_{\tau,s}(\e)$ is the density of states per unit area:
\begin{align}
    &D_{\tau,s}(\e)=
        \frac{1}{2\pi(\hbar v)^2}
        |\e-E_{\tau,s}|
        {~}\theta(|\e-E_{\tau,s}|-Z_{\tau,s}),
\end{align}
with $E_{\tau,s}=s(\tau\lambda/2-JS)$ and $Z_{\tau,s}=\Delta/2-\tau s \lambda/2$.
At zero temperature, the spin current is finite when the product of the spin-up and spin-down density of states in each valley is finite at the Fermi level as shown in Eqs. (\ref{eq_spin_current}) and (\ref{eq_susceptibility})
\cite{ominatoQuantumOscillationsGilbert2020a}

\begin{figure}
\begin{center}
\includegraphics[width=1\hsize]{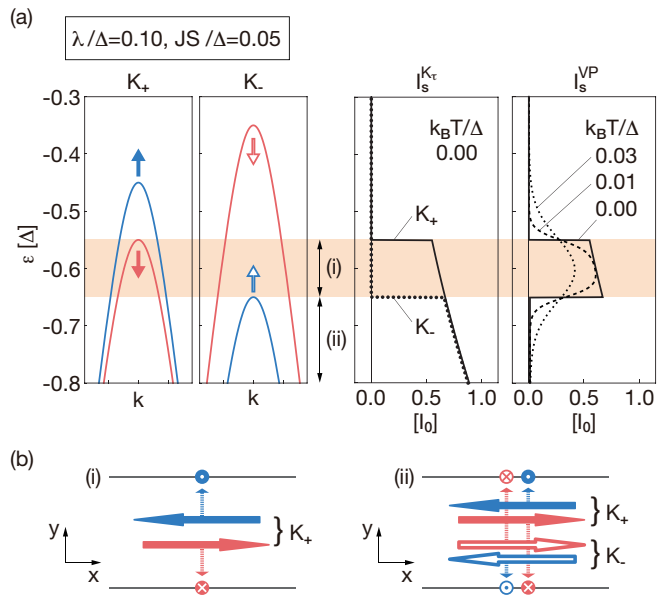}
\end{center}
\caption{
(a) The first and second panels from the left show the valence bands in the $K_+$ and $K_-$ valleys, respectively. The third panel from the left shows the spin current at the interface. The solid and dotted curves represent the valley-resolved spin current in the $K_+$ and $K_-$ valleys, respectively. The fourth panel from the left shows the valley-polarized spin current at several temperatures.
The units of the spin current are given by $I_{0}=\frac{g(\Omega)}{\pi}\frac{|J_0|^2\Delta^2}{(\hbar v)^4}\hbar\Omega$.
(b) (i) The up-spin (blue arrows) and down-spin (red arrows) electrons flowing in opposite directions lead to the transverse spin accumulation.
(ii) When the spin current is finite in the $K_+$ and $K_-$ valleys, the transverse spin accumulation cancels.
}
\label{fig_vpsc}
\end{figure}

One of the essential results of the above expressions is that the spin current can be valley polarized.
This is because the valley degeneracy is lifted in the current system so that the spin current in each valley can differ.
The valley polarization of the spin current is characterized by the valley-polarized spin current (VPSC), which is defined as
\begin{align}
    I_S^{\rm{VP}}:=I_S^{K_+}-I_S^{K_-}.
\end{align}
We show that, with appropriate carrier doping, the spin current is completely valley polarized, which means that the spins are valley-selectively excited.
The first and second panels from the left in Fig.\ \ref{fig_vpsc}(a) show the valence bands in the $K_+$ and $K_-$ valleys, respectively. The parameters are given the values $\lambda/\Delta=0.10$ and $JS/\Delta=0.05$, which are comparable to the results of first-principles calculations
\cite{qiGiantTunableValley2015a,zhangLargeSpinValleyPolarization2016,liangMagneticProximityEffect2017,xuLargeValleySplitting2018}.
The third and fourth panels from the left in Fig.\ \ref{fig_vpsc}(a) show the spin current in each valley and the VPSC, respectively.
In the energy region (i), the spin current is finite only in the $K_+$ valley, so that the spin current is completely valley-polarized.
This means that the spins are valley-selectively excited, which is feasible even at finite temperatures provided that the spin splitting due to proximity exchange coupling is much greater than the thermal broadening: $\kbt/\lambda\ll 1$.
In the energy region (ii), however, the spin current is finite in the $K_+$ and $K_-$ valleys, where the VPSC is almost zero, and the valley selectivity is suppressed.
Note also that a small spin splitting exists in the conduction band \cite{zhuGiantSpinorbitinducedSpin2011,cheiwchanchamnangijQuasiparticleBandStructure2012,kormanyosMonolayerMoSTrigonal2013,liuThreebandTightbindingModel2013}, which is omitted in the current model Hamiltonian, and valley-selective spin excitation is possible in the conduction band.

The spin current at the interface generates a diffusive spin current $\bm{j}_s$ on the TMDC monolayer (see Fig. \ref{fig_system}).
Figure \ref{fig_vpsc}(b) shows the generated diffusive spin current schematically, which consists of the flow of the electron with up-spin and the opposite flow of electron with down-spin because of the induced spin accumulation at the interface.
In the energy region (i), the diffusive spin current consists of electrons in the $K_+$ valley. Focusing on the flow of one spin, the Berry curvature leads to the transverse flow of the spin \cite{xiaoValleyContrastingPhysicsGraphene2007,yaoValleydependentOptoelectronicsInversion2008,xiaoBerryPhaseEffects2010}. 
The sign of the Berry curvature is the same for the up-spin and down-spin electrons because they belong to the same valley.
Consequently, the up-spin and down-spin electrons flowing in opposite directions lead to transverse spin accumulation, which we call the {\it spin-current Hall effect}.
This is one of the main results of this paper.
In the energy region (ii), however, the diffusive spin current consists of electrons in the $K_+$ and $K_-$ valleys.
The Berry curvatures around the $K_+$ and $K_-$ valleys have opposite signs because they are time-reversed with respect to each other.
Consequently, the transverse spin accumulations originating from the $K_+$ and $K_-$ valleys cancel each other.

\begin{figure}[t]
\begin{center}
\includegraphics[width=1\hsize]{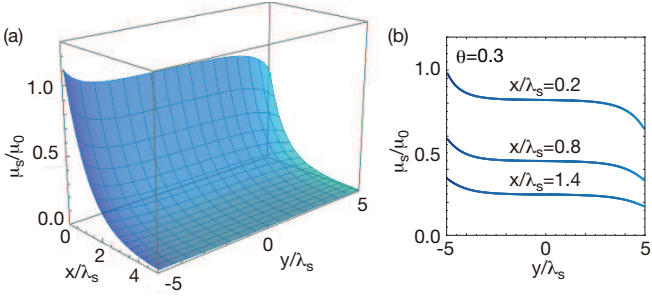}
\end{center}
\caption{
Numerical solution of spin diffusion equation with system size $L_x/\lambda_s=L_y/\lambda_s=5$ and Hall angle $\theta=0.3$.
(a) Spin accumulation $\mu_s$ plotted as a function of $x$ and $y$ and
(b) as a function of $y$ for several values of $x$.
The units of the spin accumulation are given by $\mu_{0}=ej_{0}\lambda_s/\sigma_{xx}$.
}
\label{fig_spin_diffusion}
\end{figure}

{\it Diffusion of injected spins.---}A prominent feature of the VPSC is the generation of transverse spin accumulation, as discussed above.
Here, we solve the spin diffusion equation for the valley-polarized spins to clarify the transverse spin accumulation.
The spin diffusion equation is
\begin{align}
    \left(\partial_x^2+\partial_y^2-1/\lambda_s^2\right)\mu_s(x,y)=0,
    \label{eq_spin_diffusion}
\end{align}
where $\mu_s(x,y)$ is the spin accumulation and $\lambda_s$ is the spin diffusion length.
The diffusive spin current is given by
\begin{align}
    \bm{j}_s(x,y)
        &=-\frac{\sigma_{xx}}{e}
            \left[
                \nabla\mu_s(x,y)+\theta\nabla\mu_s(x,y)\times\bm{e}_z
            \right],
    \label{eq_diffusive_spin_current}
\end{align}
where $\theta=\sigma_{xy}/\sigma_{xx}$ is the Hall angle, $\sigma_{xx}$ is the longitudinal conductivity, and $\sigma_{xy}$ is the Hall conductivity originating from the Berry curvature.
The second term describes the {\it spin-current Hall effect}.
We consider a TMDC monolayer with system size $L_x\times 2L_y$ and boundary conditions
$j_{s}^x(0,y)=j_{0},{~}j_{s}^x(L_x,y)=0$ and $j_{s}^y(x,\pm L_y)=0$.
The boundary conditions mean that the diffusive spin current is injected at $x=0$ and vanishes at the other boundaries.
We numerically solve Eq.\ (\ref{eq_spin_diffusion}) and
set $L_x/\lambda_s=L_y/\lambda_s=5$ with the parameter $\theta=0.3$ \cite{wuIntrinsicValleyHall2019,hungDirectObservationValleycoupled2019a}.
Figure \ref{fig_spin_diffusion} shows the spin accumulation decaying exponentially with the spin diffusion length in the $x$ direction, which is the usual spin diffusion.
Note the transverse spin accumulation near the boundaries $y=\pm L_y$ with opposite signs, which is the consequence of the anomalous $\theta$ term in Eq.\ (\ref{eq_diffusive_spin_current}).
In addition, the transverse spin accumulation decays with the spin diffusion length.

\begin{figure}[t]
\begin{center}
\includegraphics[width=1\hsize]{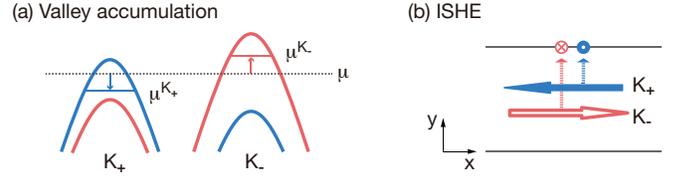}
\end{center}
\caption
{(a) Valley accumulation $\mu_v=\mu^{K_+}-\mu^{K_-}$ is induced by spin excitation with the intervalley spin-transfer processes. The dotted line represents the chemical potential $\mu$ in equilibrium.
(b) Valley accumulation with spin-valley locking leads to the inverse spin Hall effect.}
\label{fig_ISHE}
\end{figure}

{\it Discussion.---}We discuss the experimental detection of the transverse spin accumulation.
One feasible experimental technique to detect such spin accumulation is to measure the magneto-optical Kerr effect
\cite{katoObservationSpinHall2004a,sihSpatialImagingSpin2005,leeElectricalControlValley2016a}.
A spatial resolution image of the Kerr angle provides information on the spatial distribution of the spin accumulation.
Although spin-orbit coupling is strong in the TMDC monolayer, the spin diffusion length for the out-of-plane component is expected to be quite long because of the symmetry of the crystal structure. Therefore, the spin diffusion length could be comparable to the limit of the spatial resolution of the Kerr measurement (about one micron).

We also discuss the effects of intervalley spin-transfer processes at the interface, which were neglected in our main analysis.
In the presence of atomic scale interface roughness, the intervalley spin-transfer processes are not negligible and give a correction term for the spin current at the interface.
Assuming the matrix elements $J_{\bq,\bk}$ for the intervalley spin-transfer processes are approximated as a constant $J_1$, the correction term $\delta\la\IS\ra$ is estimated as $\delta\la\IS\ra\propto\sum_{\tau}|J_1|^2D_{\tau,+}(\e)D_{-\tau,-}(\e)$ because the spin current can be calculated in a similar manner performed in our main analysis.
The correction term is proportional to the product of the density of states for spin-up and spin-down electrons in the different valleys.
An vital consequence of the correction term is that non-equilibrium valley accumulation is induced by spin excitations, as shown in Fig. \ref{fig_ISHE}(a), in the presence of the spin-valley locking, which means a one-to-one correspondence between spin and valley indices at the Fermi level.

Two ways are possible to detect the valley accumulation induced by the spin excitation.
First, the valley accumulation may be detected by the inverse spin Hall effect (ISHE).
Valley accumulation leads to a diffusive spin current consisting of spin-up $K_+$ valley electrons and spin-down $K_-$ valley electrons. The valley-contrasting Berry curvature gives rise to the ISHE, which may be detected electrically.
Figure \ref{fig_ISHE} (b) shows a schematic illustration of the ISHE.
Second, valley accumulation may be detected by the modulation of the anomalous Hall conductivity, as demonstrated by the optical pumping of valley-polarized carriers \cite{makValleyHallEffect2014,ubrigMicroscopicOriginValley2017}.

Finally, we mention the candidate materials, whrere the spin-valley coupled transport phenomena by a spin excitation could be observed. In recent experimental studies, the TMDC/FI heterostructures, such as ${\rm WSe}_2$/${\rm EuS}$ \cite{zhaoEnhancedValleySplitting2017,nordenGiantValleySplitting2019} and ${\rm WSe}_2$/${\rm CrI}_3$ \cite{zhongVanWaalsEngineering2017,seylerValleyManipulationOptically2018}, are fabricated and lifting the valley degeneracy is observed in these materials, so that our theoretical predictions can be tested in these materials with an appropriate experimental setup.

{\it Conclusion.---}We present herein a study of the valley-dependent spin transport at the interface between a TMDC monolayer and a FI.
Given appropriate carrier doping, the spins are valley-selectively excited, which generates valley-polarized spin current (VPSC).
We also study spin diffusion in the TMDC. A prominent feature of the VPSC is the generation of transverse spin accumulation, which we call the {\it spin-current Hall effect}.
This valley-dependent spin excitation and transport phenomenon can expand the possibility of the future nanotechnology, and can be useful for all spin-valley logic devices
\cite{behin-aeinProposalAllspinLogic2010a}.

We thank R. Ohshima and M. Shiraishi for helpful discussions.
This work is partially supported by the Priority Program of Chinese Academy of Sciences, Grant No. XDB28000000.

\bibliographystyle{apsrev4-1}
\bibliography{../reference}

\end{document}